\documentclass{article}
\usepackage{epsfig}

\def\be{\begin{equation}}
\def\ee{\end{equation}}
\def\bea{\begin{eqnarray}}
\def\eea{\end{eqnarray}}
\def\ket#1{\hbox{$\vert #1\rangle$}}   
\def\bra#1{\hbox{$\langle #1\vert$}}   
\def\oneh{{\textstyle {1\over 2}}}
\def\onet{{\textstyle {1\over 3}}}

\def\Re{\hbox{\rm Re\,}}
\def\Im{\hbox{\rm Im\,}}

\def          
\simeq{{\ \lower2pt\hbox{$-$}\mkern-13mu \raise2pt \hbox{$\sim$}\ }} 
\def\dirac#1{\slash \mkern-10mu #1}

\begin{document}










\begin{center}
\Large{\bf Helicity-dependent generalized parton distributions in constituent
quark models}
\end{center}

\bigskip

\begin{center}
\large S.~Boffi$^a$, B.~Pasquini$^{a,b}$, M.~Traini$^c$
\end{center}

\medskip

\begin{center}
$^a$ {Dipartimento di Fisica Nucleare e Teorica, Universit\`a degli
Studi di Pavia and INFN, Sezione di Pavia, Pavia, Italy}
\end{center}
\begin{center}
$^b$ {ECT$^*$, Villazzano (Trento), Italy}
\end{center}
\begin{center}
$^c$ {Dipartimento di Fisica, Universit\`a degli Studi di Trento, Povo
(Trento), and INFN, Gruppo Collegato di Trento, Trento, Italy}
\end{center}

\bigskip

\begin{abstract}
Helicity-dependent generalized parton distributions of the nucleon are derived 
from the overlap representation of generalized parton distributions using
light-cone wave functions obtained in constituent quark models. Results from two
different quark models are used also to study the angular momentum sum rule and
the spin asymmetry in polarized electron scattering.
\end{abstract}

\hspace{0.5cm}\small Key words:
generalized parton distributions, constituent quark models

\hspace{0.5cm}PACS 12.39.-x, 13.60.Fz, 13.60.Hb, 14.20.Dh

\normalsize


\section{Introduction}

Over the years hard scattering processes in the deep inelastic scattering
(DIS) regime have provided us with considerable insight into the internal
nucleon structure, i.e. into its quark-gluon substructure as described 
by Quantum Chromodynamics (QCD). In particular, the
spin-indepen\-dent structure functions $F_1(x)$, $F_2(x)$, $F_3(x)$ and the
spin-dependent structure functions $g_1(x)$, $g_2(x)$ have been extracted as a
function of the fraction $x$ of the quark momentum along the direction of the
fast moving nucleon (for a collection of data and fitting programmes, see
Ref.~\cite{durham}). As a consequence, it has been observed that the total
nucleon momentum and spin are not exhausted by the quark contribution alone, and
a large debate has originated on the involved quark-gluon dynamics (for a recent
review, see Ref.~\cite{filippone}).

A significant step further in the investigation of such a
dynamics is offered by the recently proposed generalized parton distributions
(GPDs)~\cite{muller,radyushkin96,ji78,radyushkin97,ji55,collins97}. Their
importance is due to the fact that they are candidate to give the most
complete information on the internal nucleon dynamics and to provide us with a
unifying theoretical background suitable to describe a variety of inclusive and
exclusive processes. GPDs are non-diagonal, off-forward hadronic matrix elements of
bilocal products of the light-front quark and gluon field operators. As such
they carry information on both the longitudinal and transverse distribution of
partons in a fast moving nucleon and depend on $x$, the invariant
momentum square $t$, and the so-called skewness parameter $\xi$ describing the
longitudinal change of the nucleon 
momentum. In total, both for quarks and gluons there are four distributions
conserving the parton helicity and other four flipping it~\cite{diehl01}. In
the case of quarks, in the forward limit GPDs become diagonal matrix elements
and three of them reduce to normal quark distributions, i.e. the momentum
distribution $q(x)$, the helicity distribution $\Delta q(x)$ and the
transversity distribution $\delta q(x)$. 
Integrating the quark helicity conserving GPDs over $x$ one obtains the nucleon
electroweak form factors, that are given in terms of off-forward matrix elements
of local operators as measured in exclusive reactions. The second moment of the
unpolarized helicity conserving GDPs at $t=0$ gives a sum rule relating the
total quark contribution (including quark orbital angular momentum) to the
nucleon spin~\cite{ji78}. 

GPDs can be probed in deeply virtual Compton scattering (DVCS) and hard
exclusive production of vector mesons (for recent reviews,
see~\cite{jig,radyushkin01,goeke,marc2002,markusthesis} and references therein).
First data has become available~\cite{experiments} making the quest for modeling
GPDs more urgent.  

In the literature there are two approaches used to model the nucleon GPDs. 
One is a phenomenological construction based on reduction formulae where GPDs
are related to the usual parton distributions by factorizing the momentum
transfer dependence in agreement with the nucleon electroweak form
factors~\cite{radyushkin97,radyushkin99,radyushkin01,marc}. This leads to a
parameterization of GPDs in terms of double distribution functions. However,
this approximation is only expected to hold at relatively small values of
$-t$~\cite{penttinen}. In addition
care must be taken to add the so-called D-term~\cite{weiss}, an odd function
having support $\vert x\vert\le\vert\xi\vert$ and required to satisfy the
polynomiality property. This important property follows from Hermiticity and
time-reversal, parity and Lorentz invariance and implies that the $m$-th moment
in $x$ of a GPD at $t=0$ is an even polynomial in $\xi$ of degree less than or
equal to $m$~\cite{jig}. 

Another approach is based on direct calculation of GPDs in specific dynamical
models. The first model calculations were performed using the MIT bag
model~\cite{bag}. The four helicity conserving GPDs were studied and shown to
have a quite weak $\xi$ dependence, while their $t$ dependence roughly follows
the nucleon form factors behaviour, thus confirming the intuition at the basis
of the double-distribution assumption. However, the bag model breaks chiral
symmetry by boundary conditions at the surface, and the initial
and final nucleons are not good momentum eigenstates. As a consequence, a
support violation occurs (with GPDs small, but nonvanishing beyond $x=1$) and
the GPDs behaviour in the region $\vert x\vert\le\vert\xi\vert$ is not fully
reliable. Moreover, an antiquark distribution with negative sign is 
produced when putting three valence quarks in the bag.

Further calculations have been performed in the chiral quark-soliton
model \cite{petrov,penttinen}. The model is based on an effective relativistic
quantum field theory which was derived from the instanton model of QCD
vacuum (see Ref.~\cite{chiqsm} and references therein).
The instanton fluctuations of the gluon field are simulated by a pion field
binding the constituent quarks inside the nucleon. The model is theoretically
justified in the limit of the large number of colours $N_c$ and
satisfies all general QCD requirements (sum rules, positivity,
inequalities, etc.) including the polynomiality property~\cite{polynomiality}.
However, the model comes with an intrinsic ultraviolet cutoff in the form of a
momentum dependence of the constituent quark mass $M(p)$. The effects of the
rapidly falling function $M(p)$ are taken into account by using some
regularization scheme. The uncertainty related to the details of the ultraviolet
regularization leads to a 10--15\% numerical uncertainty of the
results~\cite{diakonov}.
In the limit of a large number of colours it describes a large variety of
nucleonic properties typically within 30\% accuracy~\cite{christov,goeke}. 
Concerning GPDs, in the leading order in the $1/N_c$ expansion it is not
possible to obtain results for separate flavours, only special flavour
combinations being nonzero, i.e. the flavour singlet part of $H(x,\xi,t)$ 
and the flavour isovector part of the unpolarized $E(x,\xi,t)$ and of the
helicity dependent GPDs, $\tilde H(x,\xi,t)$ and $\tilde E(x,\xi,t)$.
 
A complete and exact overlap representation of GPDs has been recently worked out
within the framework of light-cone quantization~\cite{diehl,brodsky}. In a
preliminary investigation of such an approach presented in Ref.~\cite{diehl99} a
good description of parton distributions at large $x$ has been achieved with a
simple ansatz for the wave functions of the three lowest Fock states. In
particular, for $x\ge 0.6$ the 80\% contribution is produced by the three
valence quarks. 

The same approach has been followed recently~\cite{BPT03}
investigating the link between light-cone wave functions (LCWFs) building the
overlap representation of GPDs and wave functions derived in constituent quark
models (CQMs). CQM wave functions can be considered as eigenfunctions of the
nucleon Hamiltonian in the instant-form dynamics and can simply be related to
wave functions in any form of relativistic Hamiltonian dynamics~\cite{keister}
according to the Bakamjian-Thomas construction~\cite{BT}. Of course, this link
is useful in the kinematic range where only (valence) quark degrees of freedom
are effective. However, in this region GPDs are obtained in a covariant approach
and exhibit the exact forward limit reproducing the parton distribution with the
correct support and automatically fulfilling the particle number and momentum
sum rule. 

The method of Ref.~\cite{BPT03} was applied to study quark helicity independent
(unpolarized) GPDs. In this paper we extend it to quark helicity-dependent
(polarized) GPDs completing the analysis of observable GPDs. In fact, both type
occurs in DVCS, while hard meson electroproduction is sensitive to unpolarized
or polarized GPDs depending on whether a longitudinal vector or a pseudoscalar
meson is produced. In contrast, to date quark helicity
flipping GPDs seem to contribute only in very peculiar selections of final
states in two vector meson electroproduction~\cite{ivanov}.

The paper is organized as follows. In Section 2 the relevant definitions are
summarized and the corresponding expressions in terms of the valence quark
contribution are given in Section 3. The results obtained within two CQMs are
presented in Section 4 where they are also applied to study nucleon spin
asymmetries occurring in inclusive scattering of polarized electrons on
polarized targets. Concluding remarks are given in the final section and 
technical details are collected in the Appendix.


\section{The helicity-dependent generalized parton distributions}

We work in the so-called symmetric frame~\cite{diehl,brodsky}. The momentum of
the initial (final) nucleon is $P^\mu$ (${P'}^\mu$). The average nucleon
momentum is then $\overline P^\mu=\oneh(P^\mu+{P'}^\mu)$. The momentum transfer
is given by $\Delta^\mu = {P'}^\mu-P^\mu$, the invariant momentum square is
$t=\Delta^2=2P\cdot\Delta$, and the so-called skewness parameter is $\xi =
-\Delta^+/2\overline P^+$. We also use the component notation $a^\mu =
[a^+,a^-,\vec{a}_\perp]$ for any four-vector $a^\mu$ with light-cone components
$a^\pm=(a^0\pm a^3)/\sqrt{2}$ and the transverse part $\vec{a}_\perp
=(a^1,a^2)$.

The helicity-dependent GPDs are defined starting from the
Fourier transform of the axial vector matrix element 

\be
\label{eq:definition}
\tilde F^q_{\lambda'\lambda}(\overline x,\xi,t) =
\left. \frac{1}{4\pi}\int dy^-\, e^{i\overline x\overline P^+y^-}
\bra{P',\lambda'}\overline\psi(-\oneh y)\,\dirac n\, \gamma^5
\psi(\oneh y)\ket{P,\lambda}
\right\vert_{y^+=\vec{y}_\perp=0},
\ee
where the four-vector $n$ is a lightlike vector proportional to $(1,0,0,-1)$, 
$\lambda$ ($\lambda'$) is the helicity of the initial (final) nucleon and
the quark-quark correlation function is integrated along the light-cone
distance $y^-$ at equal light-cone time ($y^+=0$) and at zero transverse
separation ($\vec{y}_\perp=0$) between the quarks. The resulting
one-dimensional Fourier integral along the light-cone distance $y^-$ is with
respect to the quark light-cone momentum $\overline k^+=\overline x\overline
P^+$. The link operator normally needed to make the
definition~(\ref{eq:definition}) gauge invariant does not appear because we also
choose the gauge $A^+=0$ and assume that one can ignore the recently discussed
transverse components of the gauge field~\cite{brodsky2002,belitsky2003}.

Following Ref.~\cite{ji78} the leading twist (twist-two) part of this amplitude
can be parametrized as 
\bea
\tilde F^q_{\lambda'\lambda}(\overline x,\xi,t) 
& = & \frac{1}{2\overline P^+}
\, \overline u(P',\lambda')\,\gamma^+\gamma^5\, u(P,\lambda) \, 
\tilde H^q(\overline x,\xi,t)
 \nonumber\\
& & \quad + \frac{1}{2\overline P^+} 
\, \overline u(P',\lambda')\,
\frac{\Delta^+ \gamma^5}{2M} \,u(P,\lambda)\,\tilde E^q(\overline x,\xi,t) ,
\label{eq:ampl}
\\ \nonumber
\eea
where $u(P,\lambda)$ is the nucleon Dirac spinor and $\tilde H^q(\overline
x,\xi,t)$ and  $\tilde E^q(\overline x,\xi,t)$ are the helicity-dependent GPDs
for partons of flavor $q$, corresponding on the nucleon side to an axial-vector
and a pseudoscalar transition, respectively. 

An explicit expression of the helicity-dependent GPDs in term of LCWF's has
been obtained in Refs.~\cite{diehl,brodsky}. Having in mind the link between
GPDs and CQM wave functions we will restrict our discussion into the region
$\xi<\overline x<1$. In this region and in the symmetric frame
\bea
\label{eq:base}
&\tilde F^q_{\lambda'\lambda}(\overline x,\xi,t) = & 
\sum_{N,\beta}\left(\sqrt{1-\xi}\right)^{2-N} \left(\sqrt{1+\xi}\right)^{2-N}
\sum_{j=1}^N \mbox{sign} \, (\mu_j) \, \delta_{s_jq}
\nonumber \\
& & \quad \times
\int[d\overline x]_N[d\vec{k}_\perp]_N\,\delta(\overline x-\overline x_j)
\Psi^{*}_{\lambda',N,\beta}(r') \Psi_{\lambda,N,\beta}(r)\Theta(\overline x_j),
\\ \nonumber
\eea
where $s_j$ labels the quantum numbers of the $j$-th parton, $\beta$ specifies
all other quantum numbers necessary for the $N$-parton state, and $\mu_j$ is the 
helicity of the active quark. The set of kinematical
variables $r,r'$ are defined as follows: for the final struck quark,
\be
\label{eq:finalstruck}
y'_j = \frac{\overline k^+_j + \oneh\Delta^+}{\overline P^+ + \oneh\Delta^+}
= \frac{\displaystyle \overline x_j-\xi}{\displaystyle 1-\xi},\qquad
\vec{\kappa}'_{\perp j} = \vec{k}_{\perp j} + \oneh
\frac{\displaystyle 1-\overline x_j}{\displaystyle 1-\xi}\vec{\Delta}_\perp, 
\ee
for the final $N-1$ spectators ($i\ne j$),
\be
\label{eq:finalspect}
y'_i = \frac{\displaystyle \overline x_i}{\displaystyle 1-\xi},\qquad
\vec{\kappa}'_{\perp i}= \vec{k}_{\perp i} - \oneh
\frac{\displaystyle \overline x_i}{\displaystyle 1-\xi}\vec{\Delta}_\perp,
\ee
and for the initial struck quark
\be
\label{eq:initialstruck}
y_j = \frac{\overline k^+_j - \oneh\Delta^+}{\overline P^+ - \oneh\Delta^+}
= \frac{\displaystyle \overline x_j+\xi}{\displaystyle 1+\xi}, 
\qquad
\vec{\kappa}_{\perp j}= \vec{k}_{\perp j} - \oneh
\frac{\displaystyle 1-\overline x_j}{\displaystyle 1+\xi}\vec{\Delta}_\perp,
\ee
for the initial $N-1$ spectators ($i\ne j$),
\be
\label{eq:initialspect}
y_i = \frac{\displaystyle \overline x_i}{\displaystyle 1+\xi},
\qquad 
\vec{\kappa}_{\perp i}=\vec{k}_{\perp i} + \oneh
\frac{\displaystyle \overline x_i}{\displaystyle 1+\xi}\vec{\Delta}_\perp.
\ee
Working out the spinor products we have
\bea
\tilde F^q_{++}(\overline x,\xi,t) & = & 
-\tilde F^q_{--}(\overline x,\xi,t) \nonumber\\
\label{eq:noflip}
& = &\sqrt{1-\xi^2}\,\tilde H^q(\overline x,\xi,t) 
-\frac{\xi^2}{\sqrt{1-\xi^2}}\, \tilde E^q(\overline x,\xi,t),\\
\label{eq:flippa}
\tilde F^q_{-+}(\overline x,\xi,t) & = & 
\left[F^q_{+-}(\overline x,\xi,t) \right]^*
= \eta\, \xi \, \frac{\sqrt{t_0-t}}{2M} \, \tilde E^q(\overline x,\xi,t),
\\ \nonumber
\eea
where
\be
\eta=\frac{\Delta^1 + i\Delta^2}{\vert\vec{\Delta}_\perp\vert},
\ee
and
\be
\label{eq:minimalt}
- t_0 = \frac{4{\xi}^2M^2}{1-\xi^2}
\ee
is the minimal value for $-t$ at given $\xi$.

Using Eq.~(\ref{eq:ampl}), one can derive $\tilde H^q$ and $\tilde E^q$ 
separately from the knowledge of
$\tilde F^q_{\lambda'\lambda}$. In particular, $\tilde E^q$ is directly given 
by
Eq.~(\ref{eq:flippa}), and
\be
\tilde H^q(\overline x,\xi,t) = \frac{1}{\sqrt{1-\xi^2}}
\left[ \tilde F^q_{++}(\overline x,\xi,t)
+\frac{2M\xi}{\eta\,\sqrt{t_0-t}\sqrt{1-\xi^2}}\,
\tilde F^q_{-+}(\overline x,\xi,t)\right].
\ee


\section{The valence-quark contribution}

The valence-quark contribution to GPDs is obtained by specializing
Eq.~(\ref{eq:base}) to the case $N=3$, i.e.
\bea
\label{eq:valence}
\tilde F^q_{\lambda'\lambda}(\overline x,\xi,t) & = &
\frac{1}{\sqrt{1-\xi^2}}\sum_{\lambda_i\tau_i}
\sum_{j=1}^3 \delta_{s_jq}\, \mbox{sign}\,(\mu_j)\,
\int[d\overline x]_3[d\vec{k}_\perp]_3\,\delta(\overline x-\overline x_j)
\nonumber\\
& & \quad\times 
\Psi^{[f]\,*}_{\lambda'}(r',\{\lambda_i\},\{\tau_i\}) 
\Psi^{[f]}_\lambda(r,\{\lambda_i\},\{\tau_i\})\Theta(\overline x_j),
\label{eq:overlap3q}
\\ \nonumber
\eea
where
$\Psi^{[f]}_\lambda(r,\{\lambda_i\},\{\tau_i\})$ is
the eigenfunction of the light-front Hamiltonian of the nucleon, described as a
system of three interacting quarks. It is here obtained from the corresponding
solution $\Psi^{[c]}_\lambda (\{\vec{k}_i\},\{\lambda_i\},\{\tau_i\})$ of the
eigenvalue equation in the instant-form as described in Ref.~\cite{BPT03}.
Separating the spin-isospin component from the space part of the wave function, 
\be
\label{eq:separated}
\Psi^{[c]}_\lambda (\{\vec{k}_i\},\{\lambda_i\},\{\tau_i\})
= \psi(\vec{k}_1,\vec{k}_2,\vec{k}_3)
\Phi_{\lambda\tau}(\lambda_1,\lambda_2,\lambda_3,\tau_1,\tau_2,\tau_3) ,
\ee
we have
\bea
\label{eq:transform}
& & \Psi^{[f]}_\lambda (r,\{\lambda_i\},\{\tau_i\})
=  2(2\pi)^3\left[\frac{1}{M_0}\frac{\omega_1\omega_2\omega_3}
{\overline x_1\overline x_2\overline x_3}\right]^{1/2} 
\psi(\vec{k}_1,\vec{k}_2,\vec{k}_3) 
\nonumber\\
& & \qquad \times\sum_{\mu_1\mu_2\mu_3}
{D}^{1/2\,*}_{\mu_1\lambda_1}(R_{cf}(\vec{k}_1))
{D}^{1/2\,*}_{\mu_2\lambda_2}(R_{cf}(\vec{k}_2))
{D}^{1/2\,*}_{\mu_3\lambda_3}(R_{cf}(\vec{k}_3))
\nonumber\\
& & \qquad\ {}\times
\Phi_{\lambda\tau}(\mu_1,\mu_2,\mu_3,\tau_1,\tau_2,\tau_3) ,
\\ \nonumber
\eea
where $M_0$ is the mass of the non-interacting 3-quark system,
$\omega_i=(k^+_i+k^-_i)/\sqrt{2}$, and the Melosh rotations are given by
\bea
{D}^{1/2}_{\lambda\mu}(R_{cf}(\vec{\tilde k})) & = &
\bra{\lambda}R_{cf}(\overline x M_0,\vec{k}_\perp)\ket{\mu} \nonumber\\
& = & \bra{\lambda}
\frac{m+\overline xM_0-i\vec{\sigma}\cdot(\hat{\vec{z}}\times\vec{k}_\perp)}
{\sqrt{(m+\overline xM_0)^2+\vec{k}_\perp^2}}\ket{\mu}.
\\ \nonumber
\eea

In the limit $\Delta^\mu\to 0$, where $\overline x$ goes over to the parton
momentum fraction $x$, we have 
\be
\label{eq:unpol}
\tilde H^q(x,0,0) = \Delta q(x), 
\ee
where $\Delta q(x)$ is the polarized quark distribution of flavor $q$. 
Explicitly, the following simple expression is obtained
\bea
\Delta q(x) & = &
\sum_{\lambda_i\tau_i}\sum_{j=1}^3 \delta_{\tau_j\tau_q}\,
\mbox{sign}\,(\mu_j)\, \nonumber\\
& &\quad\times \int[d\overline x]_3[d\vec{k}_\perp]_3\, \delta(x-\overline x_j) 
\vert\Psi_\lambda^{[f]}(\{x_i\},\{\vec{k}_{\perp,i}\};\lambda_i,\tau_i\})\vert^2.
\\ \nonumber
\eea

The polarized quark distribution $\Delta q(x)$ combined with the unpolarized
singlet quark distribution $q(x)=H^q(x,0,0)$ and the unpolarized nonsinglet
quark distribution $E^q(x,0,0)$ determines the quark orbital-angular-momentum
distribution $L_q(x)$, i.e.~\cite{hjilu}
\be
L_q(x) = \oneh \{x[q(x) + e^q(x)]-\Delta q(x)\},
\label{eq:oam}
\ee
where $q(x) = H^q(x,0,0)$ and $e^q(x)=E^q(x,0,0)$.
By integrating Eq.~(\ref{eq:oam}) over $x$ one recovers the angular
momentum sum rule~\cite{ji78}
\be
J^q = \oneh \int \,dx\,x\,[q(x) + e^q(x)] = \oneh\Sigma^q + L^q,
\label{eq:oamsum}
\ee
where $J^q$ is the fraction of the nucleon angular momentum carried by a quark
of the flavour $q$, i.e. the sum of spin, 
\be
\oneh \Sigma^q = \int \, dx \,\Delta q(x),
\ee
and orbital angular momentum, 
\be
L^q = \int \, dx \,L^q(x). 
\ee 
Thus, from the knowledge of $\Delta q(x)$ and the unpolarized quark 
distributions one may infer the quark orbital angular momentum $L^q$.

Furthermore, by integrating the helicity-dependent GPDs over $\overline
x,$ for any value of the skewness $\xi$, one obtains the following relations
\be
\int_{-1}^1d\overline x \tilde H^q(\overline x,\xi,t) = G^q_A(t), \quad 
\int_{-1}^1d\overline x \tilde E^q(\overline x,\xi,t) = G^q_P(t),
\label{eq:ff_sumrule}
\ee
where $G_A^q$ and $G_P^q$ are the axial vector form factor and the induced 
pseudoscalar form factor of the quark of flavour $q$, respectively.

\begin{figure}[ht]
\begin{center}
\epsfig{file=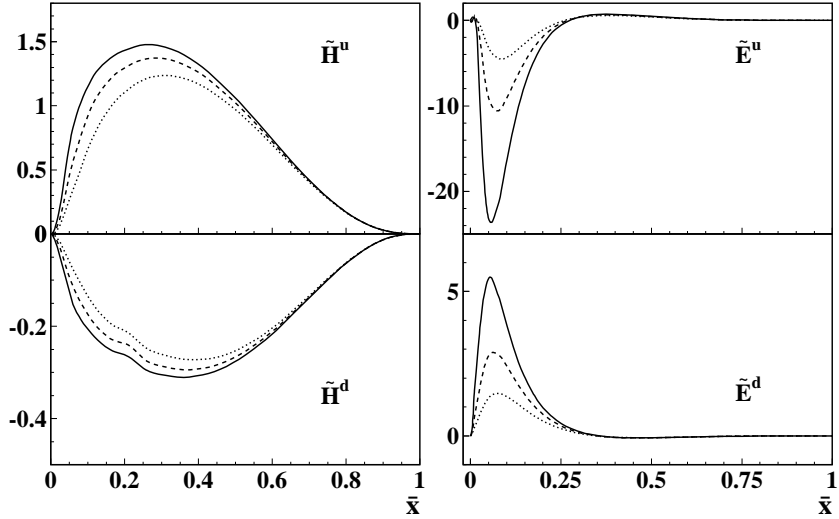,  width=26 pc}
\end{center}
\vspace{-0.4cm}
\caption{\small The helicity-dependent spin-averaged ($\tilde H^q$, left panels)
and the helicity-flip ($\tilde E^q$, right panels) generalized parton
distributions calculated in the GBE model for flavours $u$ (upper panels) and
$d$ (lower panels), at $\xi=0$ and different values of $t$: $t=0$ (solid
curves), $t=-0.2$ (GeV)$^2$ (dashed curves), $t=-0.5$ (GeV)$^2$ (dotted
curves).} 
\label{fig:fig1}
\end{figure}

\begin{figure}[ht]
\begin{center}
\epsfig{file=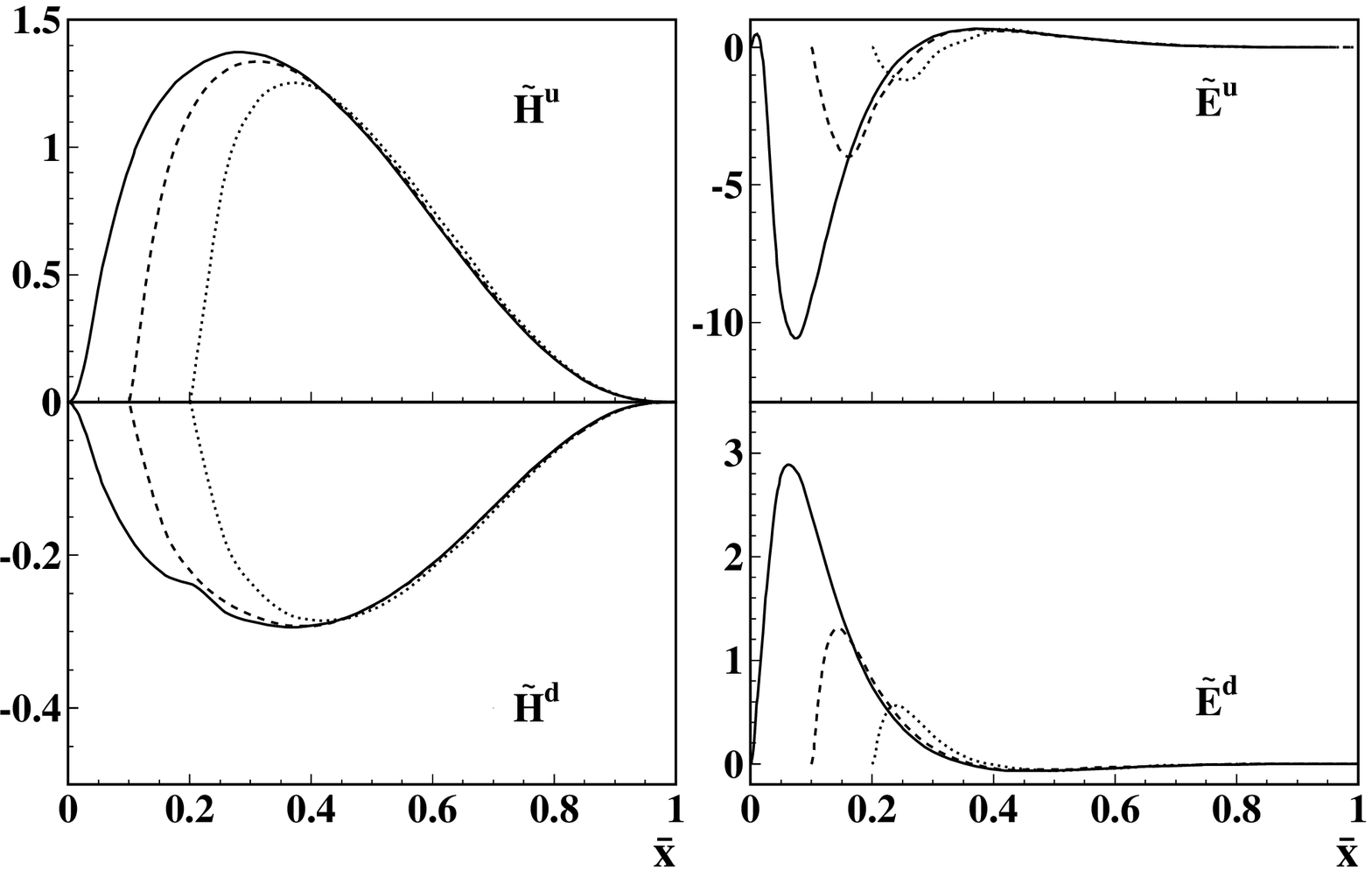,  width=26 pc}
\end{center}
\vspace{-0.4cm}
\caption{\small The same as in Fig.~\ref{fig:fig1} but for fixed $t=-0.2$
(GeV)$^2$ and different values of $\xi$: $\xi =0$ (solid curves), 
$\xi=0.1$ (dashed curves), $\xi=0.2$ (dotted curves).}
\label{fig:fig2}
\end{figure}

\begin{figure}[ht]
\begin{center}
\epsfig{file=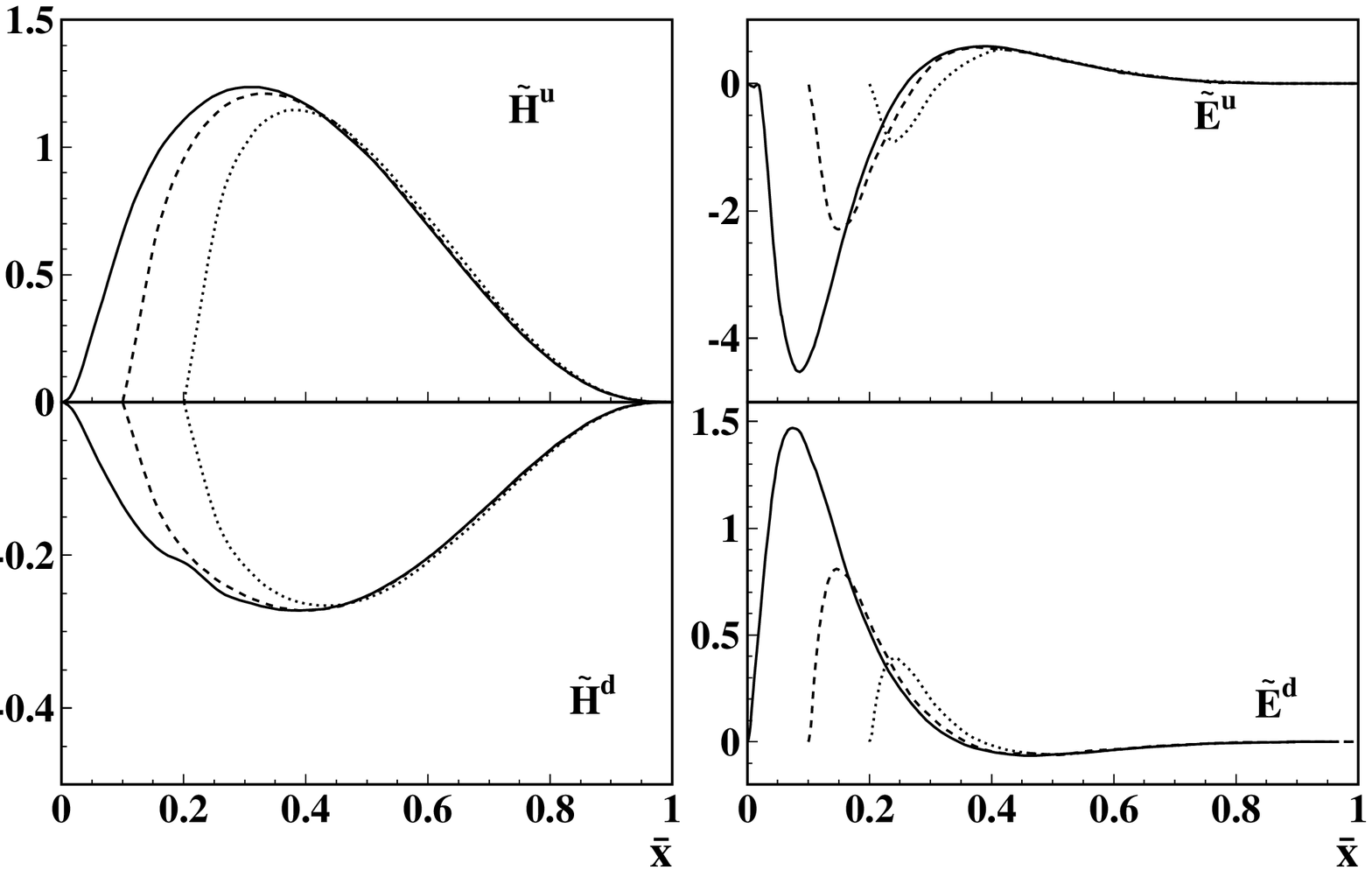,  width=26 pc}
\end{center}
\vspace{-0.4cm}
\caption{\small The same as in Fig.~\ref{fig:fig1} but for fixed $t=-0.5$
(GeV)$^2$ and different values of $\xi$: $\xi =0$ (solid curves), $\xi=0.1$
(dashed curves), $\xi=0.2$ (dotted curves).}
\label{fig:fig3}
\end{figure}


\section{Results and discussion}

In this Section we present results obtained within two constituent quark models,
i.e. the  relativistic version (HYP)~\cite{faccioli} of the hypercentral quark
model~\cite{santopinto} and the Goldstone-boson-exchange (GBE) model of
Ref.~\cite{graz}. In spite of its simplicity the hypercentral model is able
to describe the basic features of the low-lying nucleon spectrum
satisfactorily with a SU(6) symmetric nucleon wave function. With the GBE model
the baryon spectrum is well reproduced up to 2 GeV with the correct orderings of
the positive- and negative-parity states in the light and strange sectors. The
resulting nucleon wave functions, without any further parameter, yield a
remarkably consistent picture of the electroweak form factors within a
point-form approach to quark dynamics~\cite{robert}. 

Technical details concerning the derivation of the
relevant formulae with these models are given in the Appendix. 

The helicity-dependent GPDs calculated in the GBE model for the $u$ and $d$
flavours are plotted in Figs.~\ref{fig:fig1}--\ref{fig:fig3} as a function of
$\overline x$ at different values of $t$ and $\xi$. Both $\tilde H^u$ and
$\tilde H^d$ exhibit a small $t$-dependence at $\xi=0$. At constant $t$, their
(rather weak) $\xi$-dependence is entirely governed by the requirement that
$\tilde H^q$ has to vanish at the boundaries of the allowed range
$\xi\le\overline x\le 1$ as a consequence of including only valence quarks in
the present approach. Therefore the peak position of $\tilde H^u$ and $\tilde
H^d$ for increasing $\xi$ is shifted to higher values of $\overline x$. Due to
the opposite sign of $\tilde H^u$ and $\tilde H^d$ in all kinematic conditions,
their difference is positive and peaked at a value of $\overline x$ comparable
to the result obtained in the chiral quark-soliton model in the leading order of
the $1/N_c$ expansion~\cite{penttinen}. Also $\tilde E^u$ and $\tilde E^d$ have
opposite sign as functions of $\overline x$, their difference being rather
small at intermediate and large values of $\overline x$. In fact, it is known
that the difference $\tilde E^u-\tilde E^d$ is only significantly different from
zero in the region $\vert\overline x\vert\le\xi$, where it is dominated
by the contribution of the pion pole~\cite{penttinen}.

\begin{figure}[ht]
\begin{center}
\epsfig{file=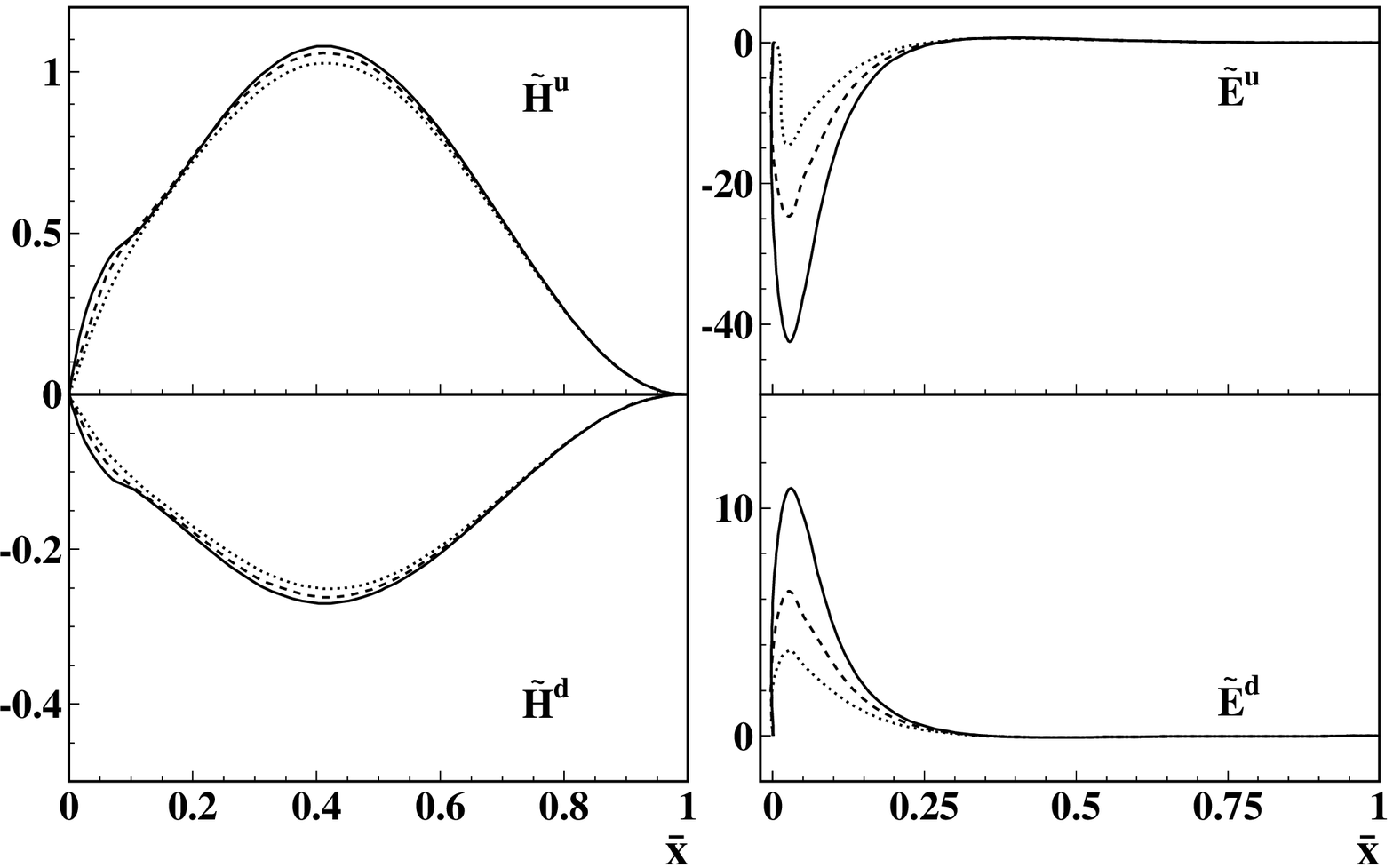,  width=26 pc}
\end{center}
\vspace{-0.4cm}
\caption{\small The same as in Fig.~\ref{fig:fig1} but for the hypercentral
model.}
\label{fig:fig4}
\end{figure}


The helicity-dependent GPDs calculated with the hypercentral and the GBE model
behave quite similarly (Fig.~\ref{fig:fig4}) in spite of the
fact that the hypercentral model is SU(6) symmetric, while the nucleon GBE wave
function contains a small SU(6)-breaking part~\cite{graz,robert}.
However one should
notice that the effect of SU(6) symmetry is not equivalent
(for spin observables in particular) to the naive idea 
suggested by nonrelativistic dynamics. Within a relativistic 
approach the correlation between motion and spin (helicity) 
and the large content of high momentum components in the 
wavefunction (determined by the relativistic kinetic
operator) change the intuitive picture considerably. In fact
the SU(6)-breaking effects are emphasized (within a relativistic approach) by
such correlations and high momentum tails, reducing the amount of explicit
SU(6)-breaking terms required by nonrelativistic approaches.
To better appreciate the symmetry effects
on GPDs, it is worthwhile to discuss first the 
integral properties of diagonal spin and angular momentum 
observables as defined in Eqs. (20)--(22).

The different contributions entering the angular momentum sum rule
(\ref{eq:oamsum}) are presented in Table~\ref{tab1} as obtained in the
nonrelativistic SU(6)-symmetric quark model, the hypercentral and the GBE model
with the same valence quark distributions derived here and in our previous
paper~\cite{BPT03}. The differences between the nonrelativistic SU(6)-symmetric
model and the hypercentral potential are entirely due to relativistic effects.
In particular, the unpolarized nonsinglet quark distribution $e^q(x) =
E^q(x,0,0)$ vanishes in the nonrelativistic limit when all the valence quark are
accommodated in the $s$-wave, while within the relativistic treatment the $u$
and $d$ contributions to $e^q$ are quite sizeable in both the hypercentral and
GBE models. It is remarkable that $u$ quarks contribute much
more than $d$ quarks. This is expected and in agreement with previous findings
from lattice QCD (see, e.g., Refs.~\cite{gockeler,haegler}). 
Only the small difference ($0.27$ in contrast to $0.20$) can be ascribed to
genuine SU(6)-breaking effects as considered by GBE. A similar behaviour can be
found for the quark spin contributions ($\Sigma^q$) and the (quark) angular
momentum components ($L^q$).

More specifically one can disentangle the effects due to Melosh 
rotations and the high momentum components including (as a 
``pedagogical'' example) Melosh rotation into the calculations of
spin observables performed with nonrelativistic wavefunctions (which
do not contain high momentum components). The effect is
somehow surprising: the total quark spin part $\Delta \Sigma$
(which reduces to $0.46$ including $u$ and $d$ components in 
the HYP model) is enhanced up to $0.75$~\cite{CAFASCOTRA00}.
Let us remark that $0.75$ (in contrast to the naive value $1.0$
ascribed to nonrelativistic quark models) is  
often quoted as the ``the relativistic reduction of quark
spin contribution to the total nucleon spin due to lower components of the
wave function''.
Our results emphasize that the actual reduction is largely
influenced by the consistent solution of the mass equation
and a simplified guess could induce large uncertainties.

\begin{table}
\caption{Valence contributions to the angular momentum sum rule calculated
within
the nonrelativistic SU(6)-symmetric quark model (NR-SU(6)), the SU(6)-symmetric
hypercentral model (HYP) and the Goldstone-boson-exchange (GBE) model.}
\label{tab1}
\begin{center}
\begin{tabular}{cccc}
\hline \\
{} & NR-SU(6) & HYP & GBE \\ \\ \hline \\
$\int dx \,x\, u(x)$ & $2/3$ & $2/3$  & $0.65$ \\
$\int dx \,x\, d(x)$ & $1/3$  & $1/3$  & $0.35$ \\
$\int dx \,x\, e^u(x)$ &  $0$    & $0.20$ & $0.27$ \\
$\int dx \,x\, e^d(x)$ & $0$     & $-0.20$ & $-0.27$ \\
$\Sigma^u$ & $4/3$  & $0.61$ & $0.79$ \\
$\Sigma^d$ & $-1/3$  & $-0.15$ & $-0.18$ \\
$L^u$ & $0$     & $0.13$  & $0.065$ \\
$L^d$ & $0$ & $0.14$ &$ 0.13$ \\
\\ \hline \\
\end{tabular}
\end{center}
\end{table}


The integral properties shown in the Table 1 share, with all the results of the
present paper, the limitation of considering valence contribution only.
A consequence is the exact cancellation of the moments $\int dx\,x\,e^u(x) +
\int dx\,x\,e^d(x)$. Indeed, the constraints of our model, i.e. no gluons, no sea and
the momentum sum rule exhausted by valence quarks only, lead to $\int
dx\,x\,[e^u(x)+e^d(x)]=0$, as shown in Table 1; a result, however, consistent
with lattice QCD calculations~\cite{haegler}.

In order to exploit all the potential features of the present approach and avoid
the limitation mentioned, we perform a Leading Order evolution~\cite{Levol}
(the only available evolution scheme established in the case of angular momentum
densities) for the moments (20)--(22). One gets~\cite{faccioli} 

\bea{1\over 2}\Delta\Sigma(Q^2) & = & {1\over 2}\Delta\Sigma(\mu^2)\; ;
\nonumber \\
L_q(Q^2) & = & (b^{-50/81} -1) {1\over 2}\Delta\Sigma(\mu^2)
+ b^{-50/81} L_q(\mu^2) - {9\over 50}(b^{-50/81} -1)\; ; 
\nonumber \\
J_g(Q^2) & = & b^{-50/81} J_g(\mu^2) - {8\over 25}  (b^{-50/81} -1)
\nonumber
\eea
where $b=\log (Q^2/\Lambda^2)/\log(\mu^2/\Lambda^2)$ 
and  $\mu^2 = 0.079$ GeV$^2$ and $\Lambda = 0.232$ GeV.

In the region $Q^2 = 1 - 10$ GeV$^2$, the hypercentral (GBE) model predictions
range from 0.04 to 0.014 (-0.03 to -0.06) for $L_q$ and from 0.20 to 0.25 for
$J_g = L_g + \Delta g$ (and both potential models), results which are in good
agreement with QCD sum rules predictions~\cite{SRJg} ($J_g \sim 0.25$) and
lattice calculations~\cite{latJg} 
($J_g = 0.20 \pm 0.07$). Of course the predictions due to the 
nonrelativistic models behave quite differently, in particular 
one would expect $L_q(Q^2) \sim -  J_g \sim -0.25$ in the range 
$Q^2 = 1-10$ Gev$^2$ since $\Delta \Sigma$ remains constant
in $Q^2$. A detailed study \cite{CAFASCOTRA00} of the quark 
angular momentum distribution confirms such a behavior showing,
in particular, a large difference of the results obtained within 
a relativistic and a nonrelativistic approach in the large-$x$
region.
In addition it is important to notice that our predictions obtained in the
framework of vanishing gluon contribution, $J_g(\mu^2) = 0$, 
would not change much considering non vanishing
$J_g(\mu^2)$ models. As a matter of fact the scale $\mu^2$ would 
rise (since at that scale the nonvanishing gluons would carry 
some momentum) and hence $b$ would be larger for a given $Q^2$.

\begin{table}
\caption{Axial and pseudoscalar coupling constants calculated with the
hypercentral (HYP) and the Goldstone-boson-exchange (GBE) model.}
\label{tab2}
\begin{center}
\begin{tabular}{ccccc}
\hline \\
{} & $g^u_A$ & $g^d_A$ & $g^u_P$ & $g^d_P$ \\
\\ \hline \\
GBE & $0.79$ & $-0.18$ & $-0.12$ & $0.04$ \\ 
HYP & $0.61$ & $-0.15$ & $-0.23$ & $0.07$ \\ \\
\hline 
\end{tabular}
\end{center}
\end{table}


\begin{figure}[ht]
\begin{center}
\epsfig{file=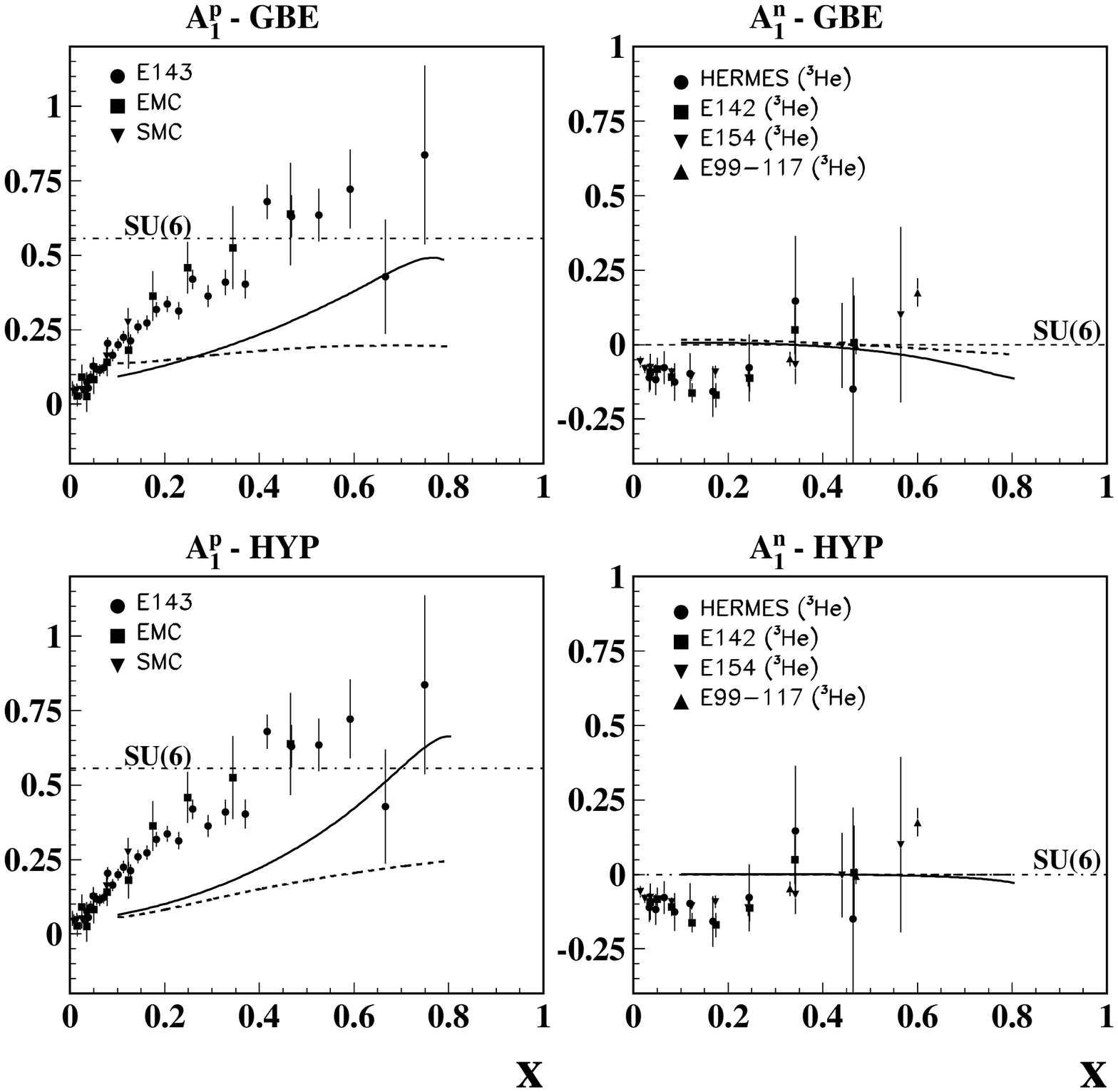, width=26 pc}
\end{center}
\vspace{-0.4cm}
\caption{\small Proton ($A^p_1$, left panel) and neutron ($A^n_1$, right panel)
spin asymmetries obtained with the GBE model (upper panels) and the hypercentral
model (lower panels). Dashed curves refer to the hadronic scale $\mu_0^2\simeq
0.1$ GeV$^2$, solid curves are the result of the evolution at $Q^2=3$ GeV$^2$ up
to NLO. Data from Refs.~\cite{e143,e155a1,emc,smca1} for
the proton and from Refs.~\cite{e142,e154g1,smcn,hermes,xiao} for the neutron.
The predicted SU(6) values are indicated by the dot-dashed line.}
\label{fig:fig7}
\end{figure}


Concerning the second moment of GPDs, Eq.~(\ref{eq:ff_sumrule}),  we
have the following axial and pseudoscalar coupling constants:
$g^{u,d}_A=G^{u,d}_A(0)$, $g^{u,d}_P=(M_\mu/2M)G^{u,d}_P(t=-0.88 M^2_\mu)$, with
$M_\mu$ being the muon mass. The calculated values are
given in Table~\ref{tab2} and are rather far from giving the experimental
values. This is evidently due to neglecting higher Fock states in the present
approach and, in the pseudoscalar case, to the missing pion-pole contribution
dominating the low-$t$ part of the induced pseudoscalar form factor.

By studying the relativistic effects introduced by the light-front
dynamics we stressed the similarities of the models we are considering,
however one could also try to investigate better the role of the SU(6)-symmetry
breaking introduced in the GBE model.
To this end one can study the spin asymmetry $A_1$. Experimentally, $A_1$ is
extracted from the ratio of polarized cross sections in the deep inelastic
regime as
\be
A_1=\frac{\sigma_{1/2}-\sigma_{3/2}}{\sigma_{1/2}+\sigma_{3/2}},
\ee
where $\sigma_{1/2(3/2)}$ is the total virtual photo-absorption cross section
for the nucleon with a projection of 1/2 (3/2) for the total spin along the
direction of the photon momentum. At high $Q^2$, $A_1$ can be approximated as
the ratio of the spin-dependent structure function $g_1$ and the
spin-independent structure function $F_2$. In addition, beyond $x=0.3$
where mainly valence quarks contribute, $A_1$ for protons and neutrons,
respectively, is given by
\be
A_1^{p}(x) = \frac{4\Delta u(x) + \Delta d(x)}{4u(x) + d(x)},\qquad 
A_1^{n}(x)= \frac{\Delta u(x) + 4\Delta d(x)}{u(x) + 4d(x)},
\ee
where $u(x)$ ($d(x)$) is the unpolarized and $\Delta u(x)$ ($\Delta d(x)$) the 
polarized quark distribution of flavour $u$ ($d$).
In the SU(6) limit with both $S=0$ and $S=1$ diquark spin states equally
contributing, one obtains $A_1^{p}=5/9$ and $A_1^{n}=0$. However, at finite
$Q^2$ one has also to consider the $Q^2$ dependence of quark distributions. This
may introduce corrections to the SU(6) limit in spite of the fact that the spin
asymmetry, being a ratio of distributions, should undergo a minimal $Q^2$
dependence.

Assuming that the valence quark distributions derived from
helicity-indepen\-dent and helicity-dependent GPDs in the approach discussed
here and in our previous paper~\cite{BPT03} correspond to the hadronic scale
$\mu_0^2\simeq 0.1$ GeV$^2$, close 
to the constituent quark mass, $Q^2$ evolution is introduced along the lines of
Ref.~\cite{mair}. Results for $A_1^{p}$ and $A_1^{n}$ obtained with the GBE and
the hypercentral model are shown in Fig.~\ref{fig:fig7}. In order to compare
with available data the hadronic-scale results have been evolved to $Q^2=3$
GeV$^2$ at next-to-leading order (NLO). As a consequence of the contribution of
sea quarks and gluons introduced by evolution, for larger $x$ the
results for the proton show an increasing trend shown also by data. In the neutron
case within CQMs it is known that a relevant role is played by SU(6)-symmetry
breaking~\cite{isgur}. Here, the very small SU(6)-breaking contribution turns
out to be in the opposite direction with respect to recent
measurements~\cite{xiao} that significantly indicates a positive $A_1^n$ at
$x=0.6$. This discrepancy can ultimately be attributed to the behaviour of the
unpolarized structure function $F_2^n$ that increases with $x$ (see also
Ref.~\cite{ratio}). Let us stress, however, that the approach here described
correctly incorporates the Pauli principle, a property crucial in
understanding SU(6)-breaking effects (see Refs.~\cite{isgur,ratio} for
discussion). 

\section{Conclusions}

The link between light-cone wave functions building the overlap representation
of generalized parton distributions and wave functions derived in constituent
quark models was introduced in Ref.~\cite{BPT03} along the lines of
Refs.~\cite{diehl,brodsky}. It has further been explored here by studying
helicity-dependent generalized parton distributions of the nucleon. The
approach, in principle correct in all kinematic regions, has an intrinsic
limitation due to the use of the lowest order Fock-space components with three
valence quark only. Results are therefore obtained in the range $\xi\le\overline
x\le 1$ and, when available, can be compared with data only in the
kinematic range where valence quarks can be assumed as the relevant degrees of
freedom. However, they can be useful, e.g., in a $Q^2$ evolution, and work
in this direction is in progress.

Two quark models have been used, i.e. the relativistic hypercentral model of
Ref.~\cite{faccioli} and the Goldstone-boson-exchange model of Ref.~\cite{graz}.
The results are in qualitative agreement with the findings of the chiral
quark-soliton model~\cite{penttinen}, a modest $t$ dependence is found for
$\tilde H^q(\overline x,\xi,t)$ at $\xi=0$, and the difference $\tilde
E^u(\overline x,\xi,t)- \tilde E^d(\overline x,\xi,t)$ is rather small in the
valence region. 

In an application to the spin asymmetry in polarized electron scattering the
same trend of data is obtained for the proton, while the opposite sign at large
$x$ is found with respect to recent data on the neutron~\cite{xiao} in contrast
with previous qualitative analyses based on CQMs with SU(6)-symmetry
breaking~\cite{isgur}; a difference which can be explained by the correct
inclusion of the Pauli principle in the present approach.

As a consequence of the correct relativistic link between wave functions of the
constituent quark model and light-cone wave functions used in the overlap
representation of generalized parton distributions, a significant contribution
of valence quarks to the quark orbital angular momentum is found. This is a
dynamical relativistic effect leading also to a nonvanishing $E(\overline
x,\xi,t)$ starting with only $s$-wave quarks, a result not obtainable in
previous approaches directly based on light-cone wave functions.



 
\appendix
\section{Appendix}
In this appendix we work out the summation over the spin and isospin variables
 appearing in the definition of the helicity amplitudes in 
 Eq.~(\ref{eq:overlap3q}).
In the case of SU(6) symmetric CQM wave functions, the summation over 
  isospin variables gives
$\delta_{T_{12}0}\,\delta_{\tau_3 1/2} +
\delta_{T_{12}1}[\delta_{\tau_3 1/2} + 2\delta_{\tau_3-1/2}]/3$ for the
proton and 
$\delta_{T_{12}0}\,\delta_{\tau_3-1/2} +
\delta_{T_{12}1}[2\delta_{\tau_3 1/2} + \delta_{\tau_3 -1/2}]/3$ for the
neutron.
On the other hand, 
the summation over the spin variables is carried out
in a similar way as in Ref.~\cite{BPT03} for the case of unpolarized GPDs, 
by using the explicit expressions of the Melosh-rotation matrices appearing in 
the initial and final light-cone wave function.
As a result, Eq.~(\ref{eq:valence}) can be rewritten as

\bea
& & \tilde F^q_{\lambda'\lambda } =
\frac{3}{2}\frac{1}{\sqrt{1-\xi^2}}\frac{1}{(16\pi^3)^2}
\int\prod_{1=1}^3 d\overline x_i\,\delta\left(1-\sum_{i=1}^3\overline x_i\right)
\,\delta(\overline x - \overline x_3) \nonumber\\
& & \qquad\times\int \prod_{i=1}^3
d^2\vec{k}_{\perp,i}\,\delta\left(\sum_{i=1}^3\vec{k}_{\perp,i}\right) \,
\tilde\psi^*(\{y'_i\},\{\vec{\kappa}'_{\perp,i}\})\,
\tilde\psi(\{y_i\},\{\vec{\kappa}_{\perp,i}\}) 
\nonumber\\
& & \qquad\times\delta_{\tau_q\tau_3}\left\{
\tilde X^{00}_{\lambda'\lambda}(\vec{\tilde{\kappa}}',\vec{\tilde{\kappa}})
\,\delta_{\tau_31/2} + \onet
\tilde X^{11}_{\lambda'\lambda}(\vec{\tilde{\kappa}}',\vec{\tilde{\kappa}})
[\delta_{\tau_31/2} + 2\delta_{\tau_3-1/2}]\right\},
\label{eq:x00_noflip}\\ \nonumber
\eea
where
\be
\tilde\psi(\{y_i\},\{\vec{\kappa}_{\perp,i}\}) =
\left[\frac{1}{M_0}
\frac{\omega_1\omega_2\omega_3}{y_1 y_2 y_3}\right]
\psi(\vec{\kappa}_1,\vec{\kappa}_2,\vec{\kappa}_3),
\ee
\bea
\tilde X^{00}_{++}(\vec{\tilde{\kappa}}',\vec{\tilde{\kappa}}) &=&
-
\tilde X^{00}_{--}(\vec{\tilde{\kappa}}',\vec{\tilde{\kappa}}) \nonumber\\
&=&
\prod_{i=1}^3 N^{-1}(\vec{\tilde{\kappa}}'_i) N^{-1}(\vec{\tilde{\kappa}}_i)
 (A_1A_2 +\vec{B}_1\cdot\vec{B}_2)\tilde A_3,
\\ \nonumber
& &\\
\tilde X^{11}_{++}(\vec{\tilde{\kappa}}',\vec{\tilde{\kappa}})& =&
-
\tilde X^{11}_{--}(\vec{\tilde{\kappa}}',\vec{\tilde{\kappa}}) 
=
\prod_{i=1}^3 N^{-1}(\vec{\tilde{\kappa}}'_i) N^{-1}(\vec{\tilde{\kappa}}_i)
\nonumber\\
& &
\times
\onet
\Big[ - (A_1A_2 +\vec{B}_1\cdot\vec{B}_2 -4B_{1,z}B_{2,z})\tilde A_3 
\nonumber\\
& &\quad+ 2 (A_1 B_{2,z}+A_2 B_{1,z})\tilde B_{3,z}\nonumber\\
& &\quad + 2(B_{1,x} B_{2,z}+B_{1,z}B_{2,x})\tilde B_{3,y}\nonumber\\
& &\quad+2(B_{1,y} B_{2,z}+B_{1,z}B_{2,y})\tilde B_{3,x}\Big],
\label{eq:x11_noflip}
\\
\Re
\, \Big(\tilde{X}^{00}_{-+}(\vec{\tilde{\kappa}}',\vec{\tilde{\kappa}})\Big)
& = &
\Re 
\, \Big(\tilde{X}^{00}_{+-}(\vec{\tilde{\kappa}}',\vec{\tilde{\kappa}})\Big) 
\nonumber\\
& = &
\prod_{i=1}^3 N^{-1}(\vec{\tilde{\kappa}}'_i) N^{-1}(\vec{\tilde{\kappa}}_i)
\Big[
(A_1A_2 +
\vec{B}_1\cdot\vec{B}_2)\tilde B_{3,y}\Big] ,\\
\Im 
\, 
\Big(\tilde{X}^{00}_{-+}(\vec{\tilde{\kappa}}',\vec{\tilde{\kappa}})\Big)
& = &-
\Im 
\, \Big(\tilde{X}^{00}_{+-}(\vec{\tilde{\kappa}}',\vec{\tilde{\kappa}})\Big) 
\nonumber\\
&=&
\prod_{i=1}^3 N^{-1}(\vec{\tilde{\kappa}}'_i) N^{-1}(\vec{\tilde{\kappa}}_i)
\Big[
 (A_1A_2 +
\vec{B}_1\cdot\vec{B}_2)\tilde B_{3,x}\Big],\\
\Re
\, \Big(\tilde{X}^{11}_{-+}(\vec{\tilde{\kappa}}',\vec{\tilde{\kappa}})\Big)
& = &
\Re 
\, \Big(\tilde{X}^{11}_{+-}(\vec{\tilde{\kappa}}',\vec{\tilde{\kappa}})\Big) 
= \prod_{i=1}^3 N^{-1}(\vec{\tilde{\kappa}}'_i) N^{-1}(\vec{\tilde{\kappa}}_i)
\nonumber\\
& &\times\onet
\Big[
(-A_1A_2 -\vec{B}_1\cdot\vec{B}_2 +4 B_{1,x}B_{2,x})\tilde B_{3,y}\nonumber\\
& &
\quad
+2(A_1 B_{2,x}+A_2 B_{1,x})\tilde B_{3,z}\nonumber\\
& &
\quad
+2(B_{1,x}B_{2,z}+B_{1,z}B_{2,x})\tilde A_3\nonumber\\
& &
\quad
+2(B_{1,x}B_{2,y}+B_{1,y}B_{2,x})\tilde B_{3,x}\Big],\\
\Im \Big(
\tilde{X}^{11}_{-+}(\vec{\tilde{\kappa}}',\vec{\tilde{\kappa}})\Big)
& = &-
\Im 
\,\Big(\tilde{X}^{11}_{+-}(\vec{\tilde{\kappa}}',\vec{\tilde{\kappa}})\Big)
=
\prod_{i=1}^3 N^{-1}(\vec{\tilde{\kappa}}'_i) N^{-1}(\vec{\tilde{\kappa}}_i)
\nonumber\\
& &
\times\onet
\Big[(-A_1A_2 -\vec{B}_1\cdot\vec{B}_2 +4 B_{1,y}B_{2,y})\tilde B_{3,x}
\nonumber\\
& &
\quad
+2(A_1 B_{2,y}+A_2 B_{1,y})\tilde B_{3,z}\nonumber\\
& &
\quad
+2(B_{1,x}B_{2,y}+B_{1,y}B_{2,x})\tilde B_{3,y}\nonumber\\
& &
\quad
+2(B_{1,y}B_{2,z}+B_{1,z}B_{2,y})\tilde A_3\Big].
\end{eqnarray}

In the above equations, $ N(\vec{\tilde{\kappa}})$,  
$A_i$ 
 and $\vec B_i,$ with $i=1,2$, are defined as in Ref.~\cite{BPT03}
and reported here for convenience
\begin{eqnarray}
N(\vec{\tilde{\kappa}}) &=& [(m+ yM_0)^2 + \vec{\kappa}^2_\perp]^{1/2}.
\\
A_i& =& (m+ y'_iM'_0)(m+ y_i  M_0) + \kappa'_{i,y} \kappa_{i,y} + \kappa'_{i,x} \kappa_{i,x},
\\
B_{i,x}& = & - (m+ y'_iM'_0) \kappa_{i,y} + (m+ y_i  M_0) \kappa'_{i,y}, 
\\
B_{i,y} &=& (m+ y'_iM'_0) \kappa_{i,x} -  (m+ y_i  M_0) \kappa'_{i,x},
\\
B_{i,z} &=& \kappa'_{i,x} \kappa_{i,y} - \kappa'_{i,y} \kappa_{i,x}, 
\end{eqnarray}

while $\tilde A_3$ and $\vec{\tilde B}_3$ are given by
\begin{eqnarray}
\tilde A_3&=&  (m+ y'_3M'_0)(m+ y_3  M_0) - \kappa'_{3,y} \kappa_{3,y} -
 \kappa'_{3,x} \kappa_{3,x},
\\
\tilde B_{3,x} &=& (m+ y'_3M'_0) \kappa_{3,y} + (m+ y_3  M_0) \kappa'_{3,y}, 
\\
\tilde B_{3,y} &=& (m+ y'_3M'_0) \kappa_{3,x} +  (m+ y_3  M_0) \kappa'_{3,x},
\\
\tilde B_{3,z} &=&  \kappa'_{3,x} \kappa_{3,y} - \kappa'_{3,y} \kappa_{3,x},
\end{eqnarray}
with $y_i,\, y'_i,$ and $\kappa_i,\,\kappa'_i,$ for $i=1,2,3,$ defined 
in Eqs.~(\ref{eq:finalstruck}) - (\ref{eq:initialspect}).

In the GBE model the nucleon wave functions are expanded on a basis where the
spin-isospin part is combined with a space part in the form of correlated
Gaussian functions of the Jacobi coordinates referring to a particular
partition. The total wave function is a symmetrized linear combination of such
basis functions over the three possible partitions thus ultimately violating
SU(6) symmetry. The calculation with the GBE wave function requires repeating
the same steps as with the hypercentral wave functions for each partition of the
partial contribution to the total initial (final) nucleon wave function. 


\clearpage
\end{document}